\documentstyle[preprint,aps,epsf,floats,colordvi]{revtex}    % PREPRINT STYLE
\def\bq{\begin{equation}}
\def\eq{\end{equation}}
\def\ba{\begin{eqnarray}}
\def\ea{\end{eqnarray}}

\tighten
\begin{document}
\thispagestyle{empty}

\newcommand{\sla}[1]{/\!\!\!#1}

\renewcommand{\small}{\normalsize} %% Remove for Phys.Rev.

\preprint{
\font\fortssbx=cmssbx10 scaled \magstep2
\hbox to \hsize{
% Next 3 lines for UW output only:
% \special{psfile=/NextLibrary/TeX/tex/inputs/uwlogo.ps
%                             hscale=8000 vscale=8000
%                              hoffset=-12 voffset=-2}
\hskip.5in \raise.1in\hbox{\fortssbx University of Wisconsin - Madison}
\hfill\vtop{\hbox{\bf MADPH-99-1106}
            \hbox{\bf FSU-HEP-990419}
            \hbox{April 1999}} }
}

\title{\vspace{.5in}
NLO QCD predictions for internal jet shapes in DIS at HERA
}
\author{N.~Kauer$^1$, L.~Reina$^2$, J.~Repond$^3$, and D.~Zeppenfeld$^1$
\\[3mm]}
\address{
$^1$Department of Physics, University of Wisconsin, Madison, WI 53706\\
$^2$Department of Physics, Florida State University, Tallahassee, FL 32306\\
$^3$Argonne National Laboratory, Argonne, IL 60439
}
\maketitle
\begin{abstract}
The transverse momentum flow inside jets is a sensitive measure 
of internal jet structure. For the current jets in deep inelastic scattering
this jet shape measure is determined at order $\alpha_s^2$, i.e. with up to 
three partons inside a single jet. The scale dependence of jet shapes in 
various jet algorithms is discussed. Results agree well with recent 
measurements by the ZEUS Collaboration, without introducing the hadronization
parameter $R_{sep}$.
\end{abstract}
%
%\pacs{PACS numbers: 13.85.-t, 14.80.Bn, 14.60.Fg}
% these PACS already chosen for ph1057
%

\newpage

%
%%%%%%%%%%%%%%%%%%%%%%%%%%%%%%%  MAIN TEXT  %%%%%%%%%%%%%%%%%%%%%%%%%%%%%%%%%
%

%\section{Introduction}\label{sec:one}

Experiments at the high energy frontier require a good understanding of 
jets, their distribution in phase space with respect to each other and to 
the leptons and photons produced in mixed electroweak-QCD processes. The 
internal structure of jets is equally important. Internal jet structure 
is intimately related to the number of jets reconstructed by different
jet-defining algorithms. It also is an issue in reconstructing the kinematic 
properties of single jets, such as their transverse momentum or direction,
and the invariant mass of two- or three-jet systems. Both
aspects are important when searching for signs of new physics in hadron 
collider experiments.

One measure of internal jet structure is the transverse energy flow 
inside jets. This internal jet shape is defined
as the fraction of a jet's transverse energy, $E_T$, which is deposited 
inside a sub-cone of radius $r<R$, where $R$ is the cone size of the jet
in the $\eta$-$\phi$ plane. Such jet shape measurements 
have been performed in the past at $p\bar p$ colliders~\cite{ua1,tevatron}
and also in photo-production and deep-inelastic-scattering (DIS) events
at HERA~\cite{h1,zeusgamma,zeusdis}, and have been compared to theoretical 
calculations~\cite{eks,kramer,giele,seymour}. Comparisons of the shape
of gluon-rich jets at hadron colliders with the quark-dominated jets produced
in $ep$ and $e^+e^-$ collisions~\cite{opal} confirm the broader structure 
of gluon jets, which is expected because the larger color charge of gluons
as compared to quarks leads to enhanced collinear radiation.

A major obstacle for a precise comparison of jet shape data to perturbative 
QCD predictions is the fact that, at tree level, QCD jets are single, 
massless partons. Hence, a nontrivial jet shape only arises 
at higher order, where a jet may contain two or more partons. Two partons 
inside a jet, which appear in typical NLO cross section calculations, 
produce jet shapes at lowest order only. Jets with three partons first appear
in two-loop calculations and thus a determination of jet shapes at true NLO, 
in a given physical process, is extremely demanding theoretically. 
Photo-production of jets~\cite{kramer} or dijet production at the 
Tevatron~\cite{eks} is a case in point: the kinematics of the event requires 
at least two hard jets in the final state, with balancing transverse momentum. 
A NLO jet shape, with jets consisting of three partons, thus requires 
four-parton final states at tree 
level and one-loop corrections to all $2\to 3$ processes involving quarks 
and gluons. These corrections are only now becoming available~\cite{nlo3jet}.

In this letter we perform a full NLO jet shape calculation in a 
kinematically simpler situation: DIS at HERA. At sufficiently large $Q^2$,
the scattered electron or positron in a DIS event provides the transverse 
momentum which is required to balance a high-$E_T$ jet. A three-parton final 
state then suffices to generate jets containing three colored partons. The 
soft and collinear divergences, which are generated by integrating the 
three-parton 
contributions over the entire phase space, are canceled by one-loop
corrections to two-parton final states and two-loop corrections to one-parton
contributions. However, a single parton cannot produce internal jet structure
and thus all true two-loop effects can be neglected when determining 
differential jet shapes for up to three partons. 
The full one-loop QCD corrections for two- and three-parton final states 
in DIS are implemented in the MEPJET Monte 
Carlo program~\cite{mepjet}. Consequently, it is possible to extract 
full NLO jet shapes for the current jet in DIS with MEPJET.

In the following, we analyze NLO corrections to the differential jet shape
$\rho(r,R,E_{Tj},\eta_j)$ for events with a single jet of transverse 
energy $E_{Tj}$ and pseudo-rapidity $\eta_j$. The differential jet shape 
is defined as 
\bq
\label{eq:rho}
\rho(r,R,E_{Tj},\eta_j) = {1\over d^2\sigma_{NLO}/dE_{Tj}\, d\eta_j}
 \sum_{n}\; \int dr_n\; {E_{Tn} \over E_{Tj}}   
\delta(r-r_n) {d^3 \sigma \over dE_{Tj}\, d\eta_j\, dr_n}\; .
\eq
Here the sum runs over all partons, $n$, belonging to the jet, which have a 
separation $r_n = \sqrt{(\eta_n-\eta_j)^2+(\phi_n-\phi_j)^2}=r<R$ from the 
axis of the jet in the legoplot. In practice, both the data and the Monte 
Carlo calculation replace the differential distributions by integrals over
finite bin sizes in $E_{Tj}$, $\eta_j$ and $r$. 
Since we want to compare our NLO results with the ZEUS data, we follow the 
ZEUS event selection~\cite{zeusdis} and study current jets in neutral 
current (NC) DIS events with
\bq
\label{eq:cutszeus}
E_e>10\;{\rm GeV}\;, \qquad Q^2>100\;{\rm GeV}^2\; , 
\qquad E_{Tj}>14\;{\rm GeV}\;, \qquad -1<\eta_j<2\; ,
\eq
where $E_e$ is the energy of the scattered electron in the laboratory frame.
The default jet cone size is $R=1$ and cone slices of width $\Delta r = 0.1$
are considered.
Because of the modest jet $E_T$, $b$-quarks are unlikely to appear inside
a current jet. Production of massive $\bar bb$ pairs via photon-gluon 
fusion is small also, for the cuts considered below. Thus it is appropriate 
to perform the calculations in a fixed 4-flavor scheme, neglecting any 
$b$-quark contributions. Matching 4-flavor parton distributions are provided 
by the CTEQ4F4 set~\cite{cteq4f4}; for consistency with the parton 
distribution functions we use two-loop expressions for the running strong 
coupling constant, with $\Lambda^{(4)}_{\overline {\rm MS}}=292\;{\rm MeV}$.

For the theoretical jet shapes, the factors in the denominator of
Eq.~(\ref{eq:rho}) are determined using NLO DIS cross 
sections at ${\cal O}(\alpha_s)$, within a given set of 
selection cuts. Therefore, they correspond to 1-jet exclusive cross 
sections for the 1-jet cuts described below.  For the ZEUS cuts, 2-jet 
events also enter with single weight only. The numerator is determined at full 
${\cal O}(\alpha_s^2)$ for the NLO results and at ${\cal O}(\alpha_s)$ 
for the LO calculations.
The default version of the MEPJET2.2 program is written for the calculation
of NLO dijet cross sections in DIS, i.e. allowing for one soft or collinear 
parton in the final state. For the calculation of jet shapes at NLO, up to 
two soft and/or collinear partons must be generated. 
We have modified the MEPJET
phase space generator to cover this enlarged phase space region for two- and 
three-parton final states. 

The matrix elements needed for the calculation of NLO jet shapes are identical
to the ones used for the NLO dijet cross section and have been tested 
previously~\cite{mepjet,graudenz}. The cancellation of collinear and infrared 
singularities employs the phase space slicing method of Giele, Glover, and 
Kosower~\cite{ggk}, which removes soft and/or collinear regions of the 
three-parton phase space, where some pair of final and/or initial parton 
momenta
satisfies $2p_i\cdot p_j < s_{min}$. Contributions from these regions are 
approximated by the appropriate asymptotic expressions and added analytically
to the two-parton contributions, where they cancel the soft and collinear 
singularities of the virtual contributions. The final result must be 
independent of the choice of $s_{min}$, for values of this soft cutoff 
sufficiently small to make the asymptotic approximations valid. 

\begin{figure}[t]
\vspace*{0.5in}
\begin{picture}(0,0)(0,0)
\includegraphics{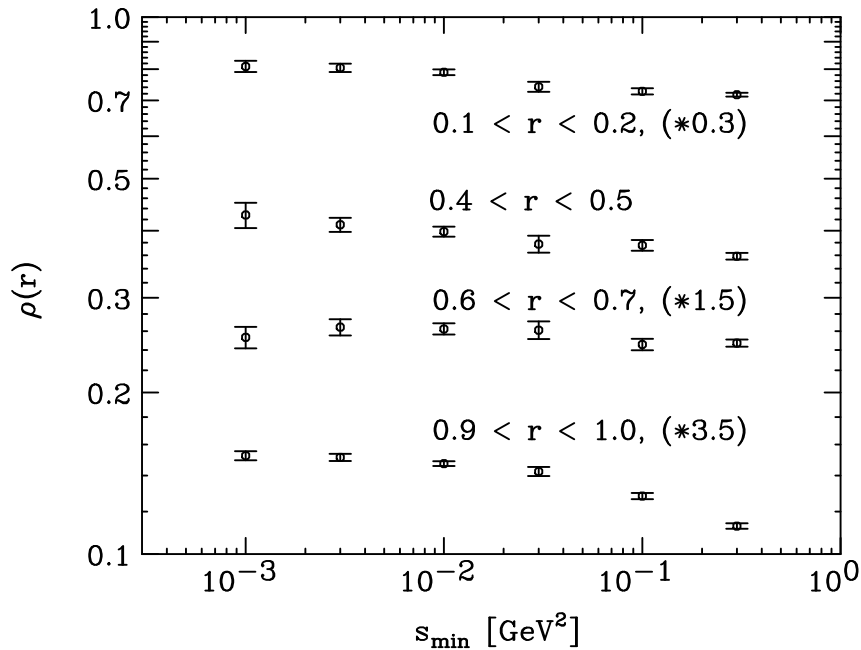}
\end{picture}
\vspace{7.0cm}
\caption{$s_{min}$ dependence of jet shapes $\rho(r)$ in four representative
$r$-bins. The Ellis-Soper $k_T$ algorithm~\protect\cite{es} has been used, 
within the cuts of Eqs.~(\protect\ref{eq:cutszeus},\protect\ref{eq:cutsopt})
and requiring that only one hard jet, of $E_{Tj}>5$~GeV is seen in the 
pseudo-rapidity range $-2.5<\eta_j<2.5$. 
Errors represent Monte Carlo statistics only. Note that results in three of
the four bins have been rescaled by the factors given in parentheses.
}
\vspace*{0.2in}
\label{fig:smin}
\end{figure}

Numerical $s_{min}$
independence is a powerful test for the correct implementation of observables
and of the phase space generator, as well as for the infrared stability of the
jet clustering algorithms~\cite{nlo3jet}.
Results of this $s_{min}$ test are shown in Fig.~\ref{fig:smin} for the 
differential jet shapes at $r=0.15$, 0.45, 0.65, and 0.95 in the successive
combination algorithm of Ellis and Soper~\cite{es} (see below). Within
Monte Carlo statistical errors of about 1-5\%, results are 
independent of $s_{min}$ below $s_{min}\approx 0.01\;{\rm GeV}^2$.
We have checked that similar
results hold for our implementation of the PUCELL jet algorithm employed by 
ZEUS~\cite{zeusgamma}. We will use $s_{min}=0.01\;{\rm GeV}^2$ in the 
following.

In our NLO calculation of jet shapes, the current jet contains up to three
partons, which provides a much more detailed simulation of internal 
jet structure than is possible in a LO analysis. The extra detail suggests 
a direct comparison of NLO theory with 
data, with the $E_T$ flow of hadrons in the data replaced by 
parton $E_T$. Previous LO analyses modeled hadronization effects 
by introducing a phenomenological parameter, $R_{sep}$, which controlled 
the clustering of partons with legoplot separations between $R$ 
and $2R$\cite{rsep}. Our analysis implements the full experimental jet 
algorithms at the parton level, without introducing extra, tunable parameters. 
We have studied two algorithms in detail, the successive 
combination algorithm of Ellis and Soper~\cite{es}, called $k_T$ algorithm 
in the following, and the PUCELL algorithm used by the ZEUS 
Collaboration~\cite{zeusgamma,zeusdis}, for default cone sizes of $R=1$.

The $k_T$ algorithm successively combines pairs of nearest partons/protojets
to new protojets, up to a distance $R$ in the legoplot. Initially all 
partons are classified as protojets of transverse energy $E_T=p_T$. The
algorithm then compares the $E_T$'s of protojets, via $d_i=E_{Ti}R$, with the 
$E_T$ weighted distances, $d_{ij}= {\rm min}(E_{Ti},E_{Tj})
\sqrt{(\eta_i-\eta_j)^2+(\phi_i-\phi_j)^2}$, of pairs of 
protojets in the legoplot.
The pair of protojets with the smallest $d_{ij}$ is recombined if 
$d_{ij}$ is smaller than all the $d_i$ (which implies that their distance 
in the legoplot is smaller than $R$). Otherwise the protojet with the smallest 
$E_T$, and hence the smallest $d_i$, is called
a jet and eliminated from the list of protojets. This process is iterated 
until all protojets have been assigned to jets. Note that some of these
jets may be eliminated by subsequent selection cuts, as in 
Eq.~(\ref{eq:cutszeus}). Some freedom exists in the assignment of jet momenta,
i.e. in the recombination scheme. For the $k_T$ algorithm we use the 
$E$-scheme, i.e. the jet four-momentum is the sum of the four-momenta of the 
partons belonging to the jet and the jet $E_T$ is the sum of the parton 
$p_T$'s.
Apart from the definition of recombined momenta, our $k_T$ algorithm is 
identical to the ones described in Refs.~\cite{seymour,es}. 

The PUCELL algorithm is an iterative fixed cone algorithm. In a first step 
all partons with $p_T>300$~MeV are considered as seeds for the formation
of pre-clusters. Starting with the highest $p_T$ seed, a pre-cluster is 
formed, containing this seed and all seeds within a cone of radius $R$ 
around it. This procedure is iterated with the remaining seeds, which do 
not yet belong to a pre-cluster, in order of decreasing $p_T$. All the seeds 
belonging to a given pre-cluster are then merged, according to the Snowmass
convention~\cite{snowmass}, i.e., the legoplot variables of the pre-clusters
are defined as the $E_T$-weighted averages over the seeds,
\ba
\label{eq:snowmass1}
E_{T,p.c.} & = & \sum_i E_{Ti} \;,\\
\label{eq:snowmass2}
\eta_{p.c.}& = & {1\over E_{T,p.c.}} \sum_i E_{Ti}\;\eta_i \;,\\
\label{eq:snowmass3}
\phi_{p.c.}& = & {1\over E_{T,p.c.}} \sum_i E_{Ti}\;\phi_i  \;.
\ea
In a second step, all partons are considered, even if they fall below the 
seed threshold of 300~MeV. A new cluster axis is calculated as in 
(\ref{eq:snowmass1}-\ref{eq:snowmass3}),
in terms of all partons inside a cone of radius $R$ of the old 
(pre)cluster axis. This second step is iterated until the contents of 
all clusters stabilize. 
A third step deals with overlapping clusters, i.e. with clusters which share 
partons. If the energy of the common partons is more than 75\% of the 
energy of the lower-energy cluster, the two clusters are merged to a single 
jet. Otherwise two separate jets are formed and common partons are 
assigned to the nearest jet. 

\begin{figure}[t]
\vspace*{0.5in}
\begin{picture}(0,0)(0,0)
\includegraphics{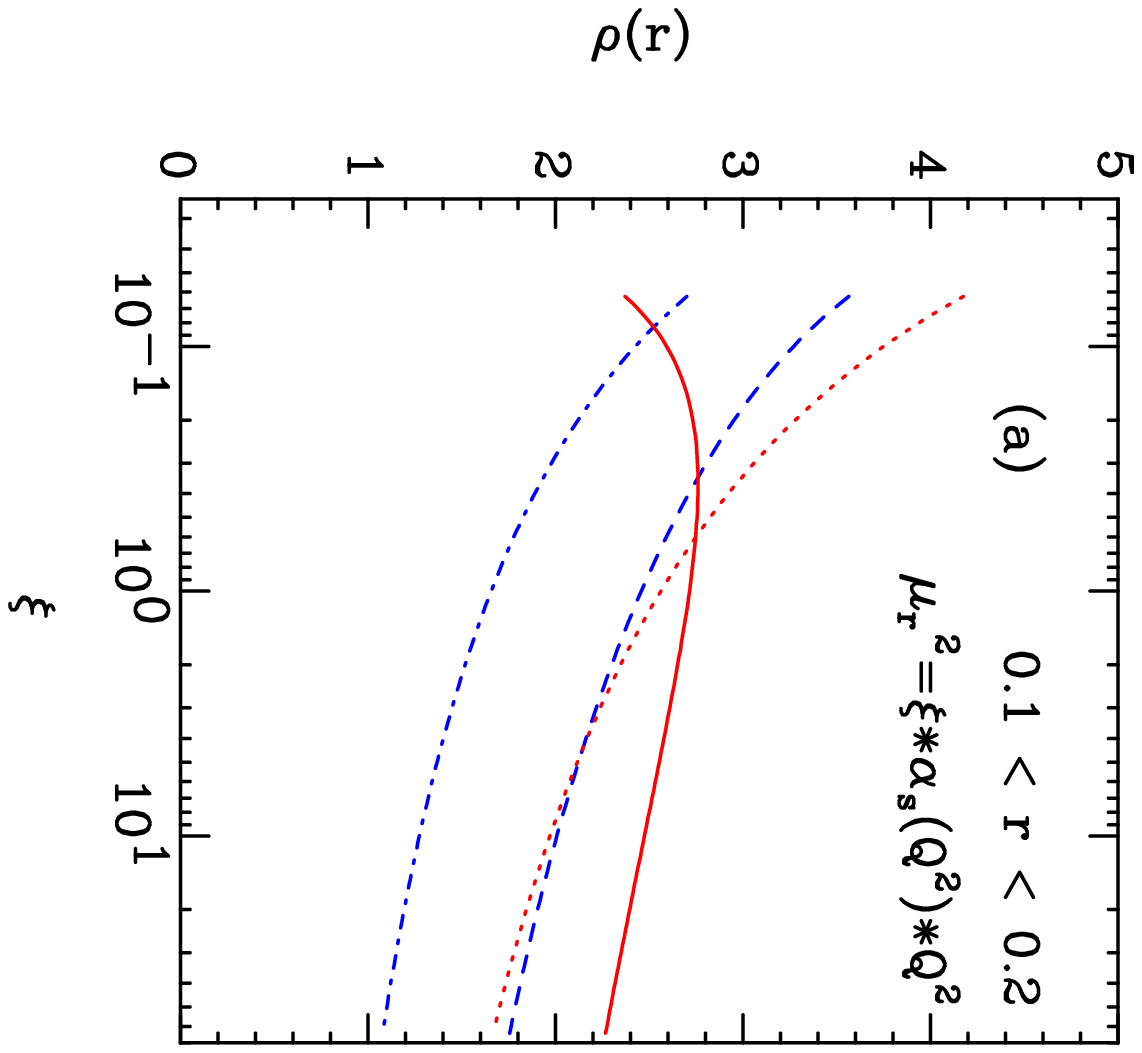}
\includegraphics{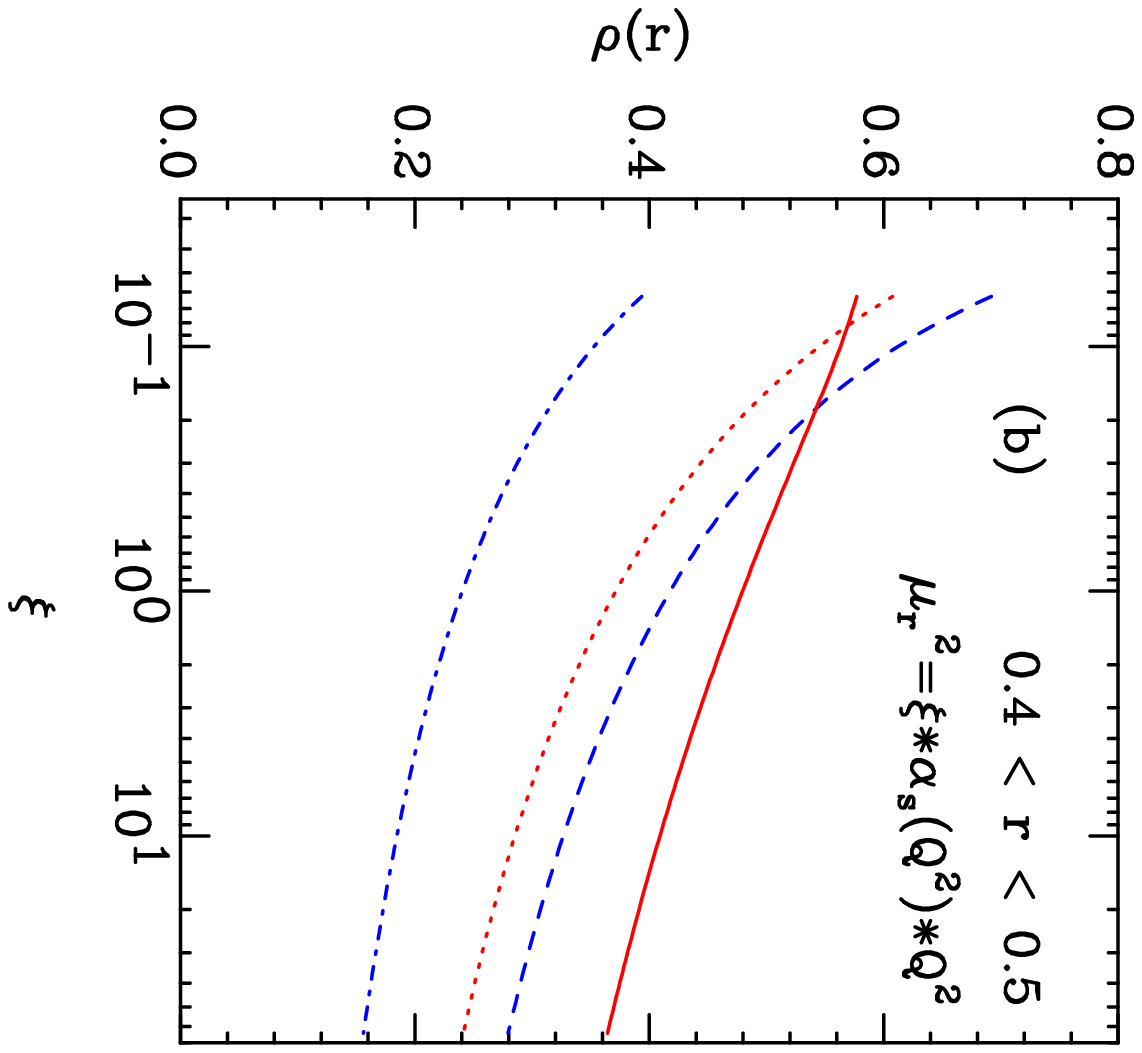}
\end{picture}
\vspace{7.0cm}
\caption{Renormalization scale dependence of the differential jet shape
$\rho(r)$ at (a) $r=0.15$ and (b) $r=0.45$, using the PUCELL algorithm. 
\Blue{The dash-dotted and dashed 
lines are the LO and NLO results for the acceptance cuts of 
Eq.~(\protect\ref{eq:cutszeus})}. \Red{Also shown are LO (dotted line) and
NLO (solid line) results for events with one single jet only of $E_T>5$~GeV
(see Eqs.~(\protect\ref{eq:jetveto},\protect\ref{eq:cutsopt}))}.
}
\vspace*{0.2in}
\label{fig:scale.PUC}
\end{figure}

We can now compare the jet shape predictions for these jet algorithms at
both leading and next-to-leading order. 
One way to assess the improvements from a NLO calculation is to study the 
scale dependence of observables. The process at hand is inclusive DIS which
basically has a single scale only, $Q$, the momentum transfer carried by 
the virtual photon. $Q$ is the obvious choice for the factorization scale 
and we fix $\mu_f^2=Q^2$ throughout. On the other hand, we are investigating 
the inner structure of jets, 
and the average jet mass or the average transverse momentum of partons,
with respect to the jet axis, appear as reasonable scale 
choices\footnote{The mass of the jet in an individual event appears as
an integration variable in the determination of $\rho(r)$ and is not 
a physical scale of the observable.}. These
scales are proportional to the intrinsic scale of the entire process, $Q$,
and become different from zero only at ${\cal O}(\alpha_s)$, thus 
suggesting $\alpha_s Q^2$ as a scale choice. 
This leads us to investigate variations of the renormalization scale
$\mu_r^2=\xi \alpha_s(Q^2)Q^2$ with the scale factor $\xi$. Searching 
for minimal sensitivity~\cite{stevenson} of $\rho(r)$, a ``correct''
scale $\mu_r=Q$ would appear as a flat $\xi$-dependence near $\xi=1/\alpha_s$.
Thus, for all practical purposes, our choice is general enough.
For the PUCELL algorithm 
and two representative bins in $r$, the scale variation of the differential
jet shape, $\rho(r)$, is shown in Fig.~\ref{fig:scale.PUC}.

The dash-dotted and dashed lines show the LO and NLO results for the 
generic ZEUS acceptance cuts of Eq.~(\ref{eq:cutszeus}). The $E_T$ flow is 
somewhat higher at NLO, i.e. jets are broader. However, the renormalization 
scale dependence is almost as large at NLO as for the LO case. This 
disappointing result can be traced to a large contribution from events with 
an additional low $E_T$ jet. In DIS events with two jets, at 
${\cal O}(\alpha_s^2)$, at best one jet can contain two partons, and, hence,
the jet shape is modeled at LO only. In order to enhance the contribution
from jets with the maximal number of partons, which then are truly modeled
at NLO, events which contain additional jets should be eliminated. 
We achieve this goal by vetoing events containing any additional jets with 
\bq
\label{eq:jetveto}
E_{Tj}>5\;{\rm GeV}\;, \qquad -2.5<\eta_j<2.5\; .
\eq
These veto cuts are chosen such that they can be easily implemented 
experimentally. A further reduction of events with hadronic activity 
beyond the current jet is achieved by requiring the scattered lepton 
and the observed hard jet to be back-to-back in azimuth. Allowing for 
finite detector resolution we require
\bq
\label{eq:cutsopt}
\Delta\phi_{ej}=|\phi_e-\phi_j|>3\;.
\eq
Also shown in Fig.~\ref{fig:scale.PUC} is the scale dependence of $\rho(r)$
for events with the jet veto cut of Eq.~(\ref{eq:jetveto}) and the 
back-to-back cut of Eq.~(\ref{eq:cutsopt}). In the following we call these 
restrictions ``1-jet cuts''. For these 1-jet events the scale independence 
of $\rho(r)$ is significantly improved at NLO, out to distances
of $r\approx 0.6$. 
  
The factors in the denominator 
of Eq.~(\ref{eq:rho}) are very stable against scale
variations for inclusive DIS events, i.e. within the ZEUS cuts of 
Eq.~(\ref{eq:cutszeus}). The 1-jet cuts, however, lead to a sizable 
reduction of $d^2\sigma_{NLO}/dE_{Tj}\, d\eta_j$ in the denominator of 
Eq.~(\ref{eq:rho}), corresponding to a subtraction of two-parton final states
with dijet-type kinematics. This subtraction term is modeled at leading 
order only and is strongly scale dependent. However, it can be 
calculated at ${\cal O} (\alpha_s^2)$ also, which results in a more 
reliable determination. 
For the 1-jet cuts we find minimal scale sensitivity~\cite{stevenson} 
near $\mu_r^2= \alpha_s(Q^2)Q^2$, and at this value the
${\cal O} (\alpha_s)$ and ${\cal O} (\alpha_s^2)$ results for the subtraction
terms virtually agree. We therefore use $\mu_r^2= \alpha_s(Q^2)Q^2$ in
the denominator, which increases the normalization of the resulting jet 
shapes by 10-15\% as compared to the choice $\mu_r=Q$.

\begin{figure}[t]
\vspace*{0.5in}
\begin{picture}(0,0)(0,0)
\includegraphics{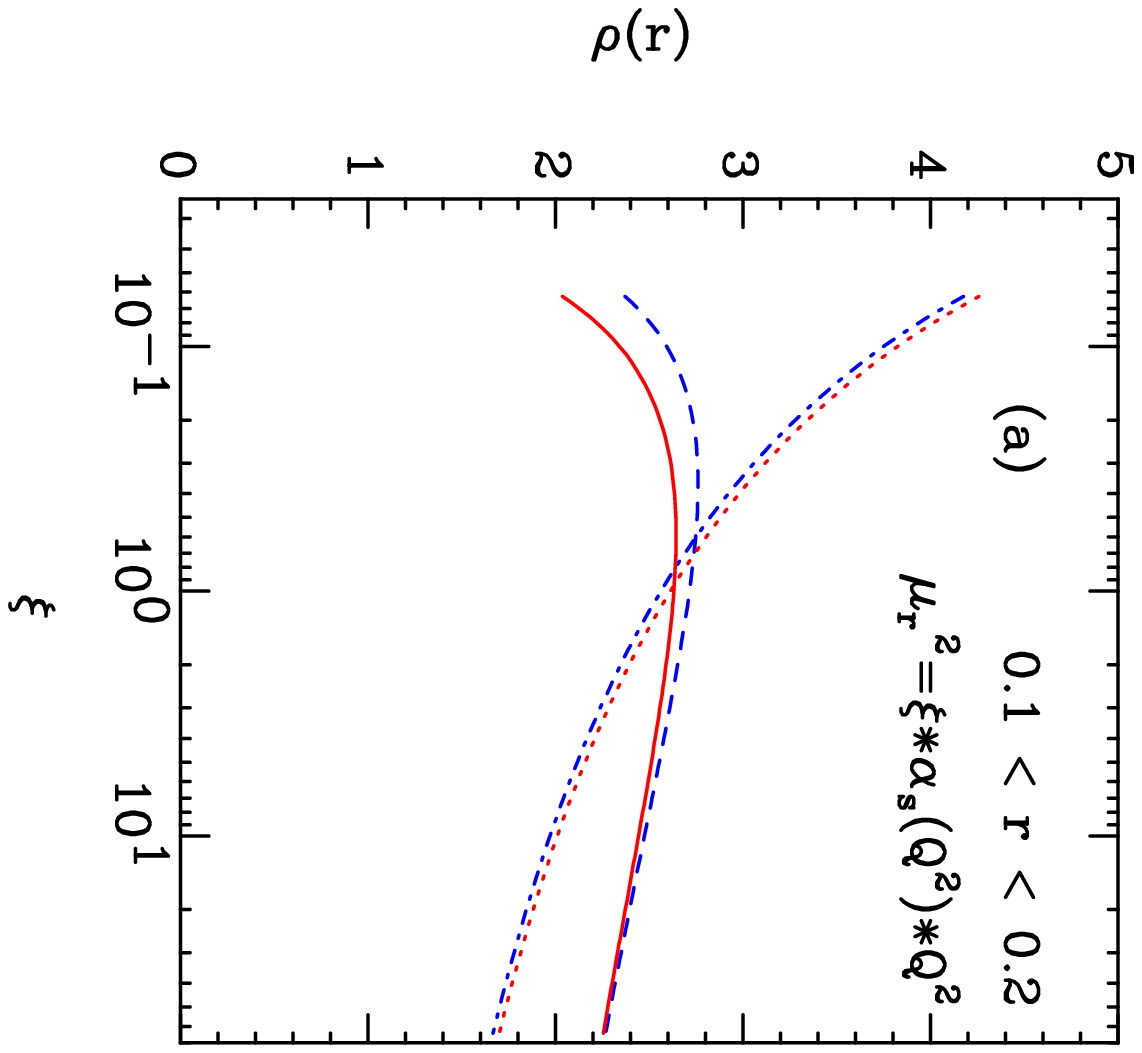}
\includegraphics{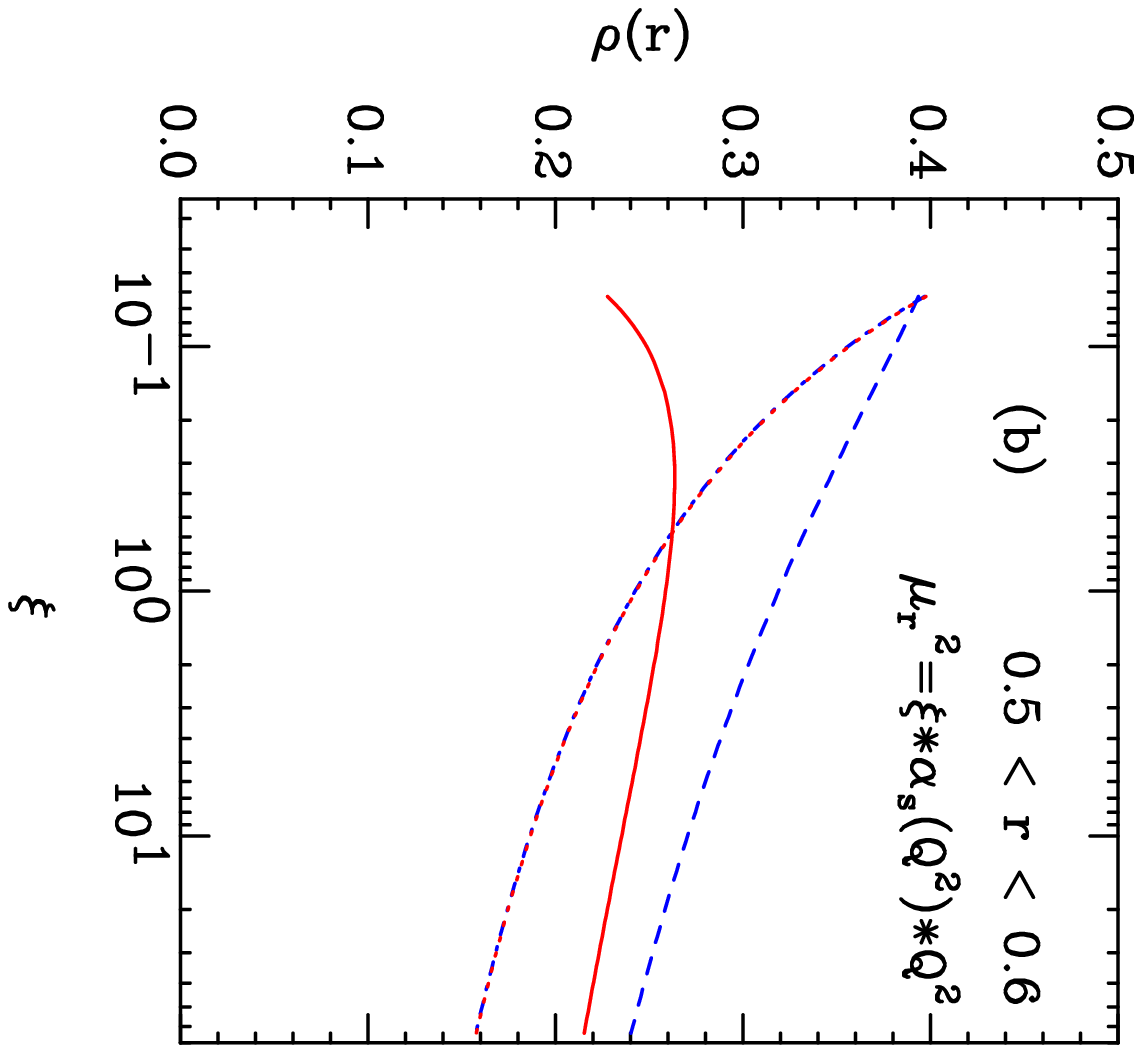}
\includegraphics{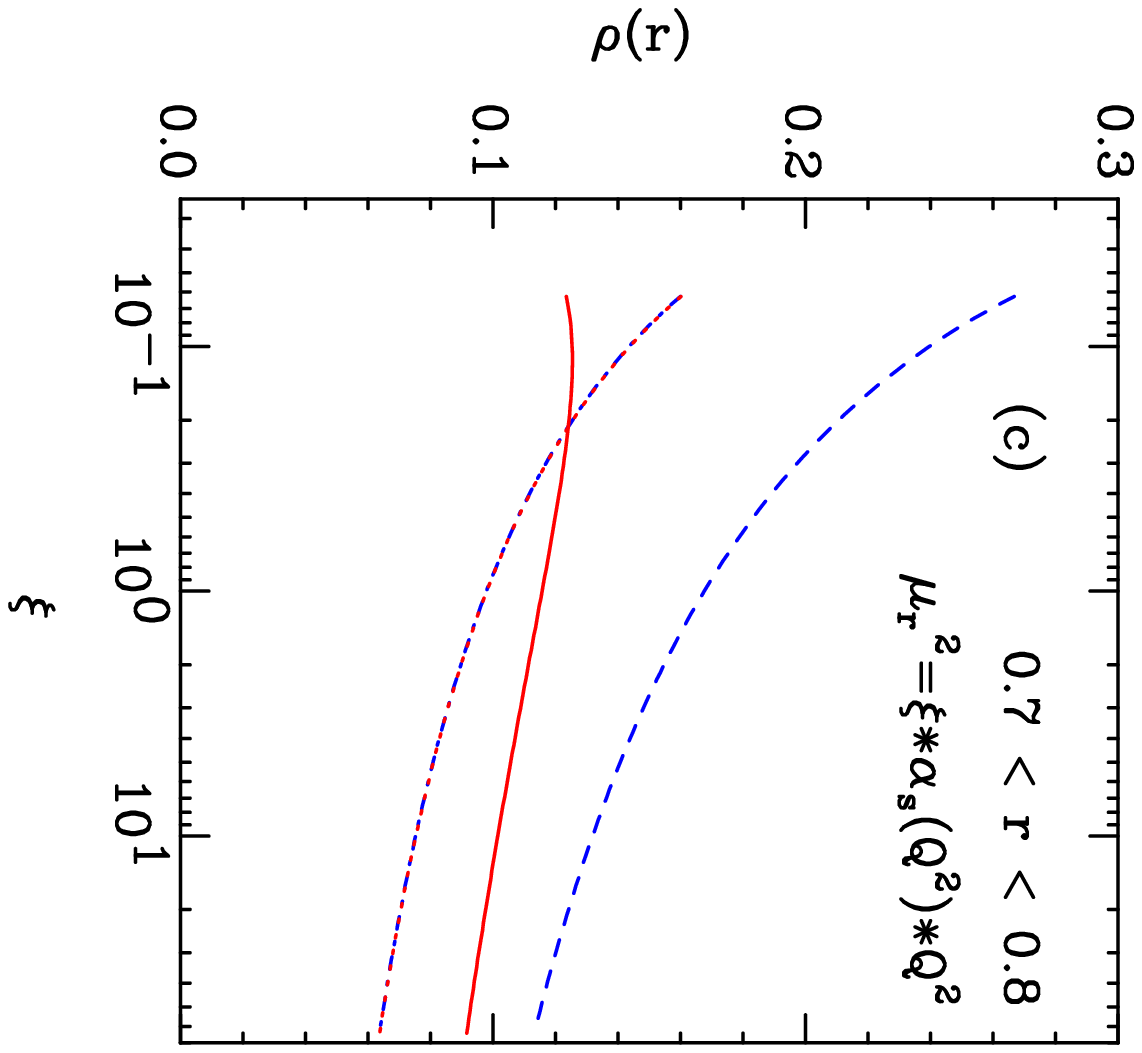}
\includegraphics{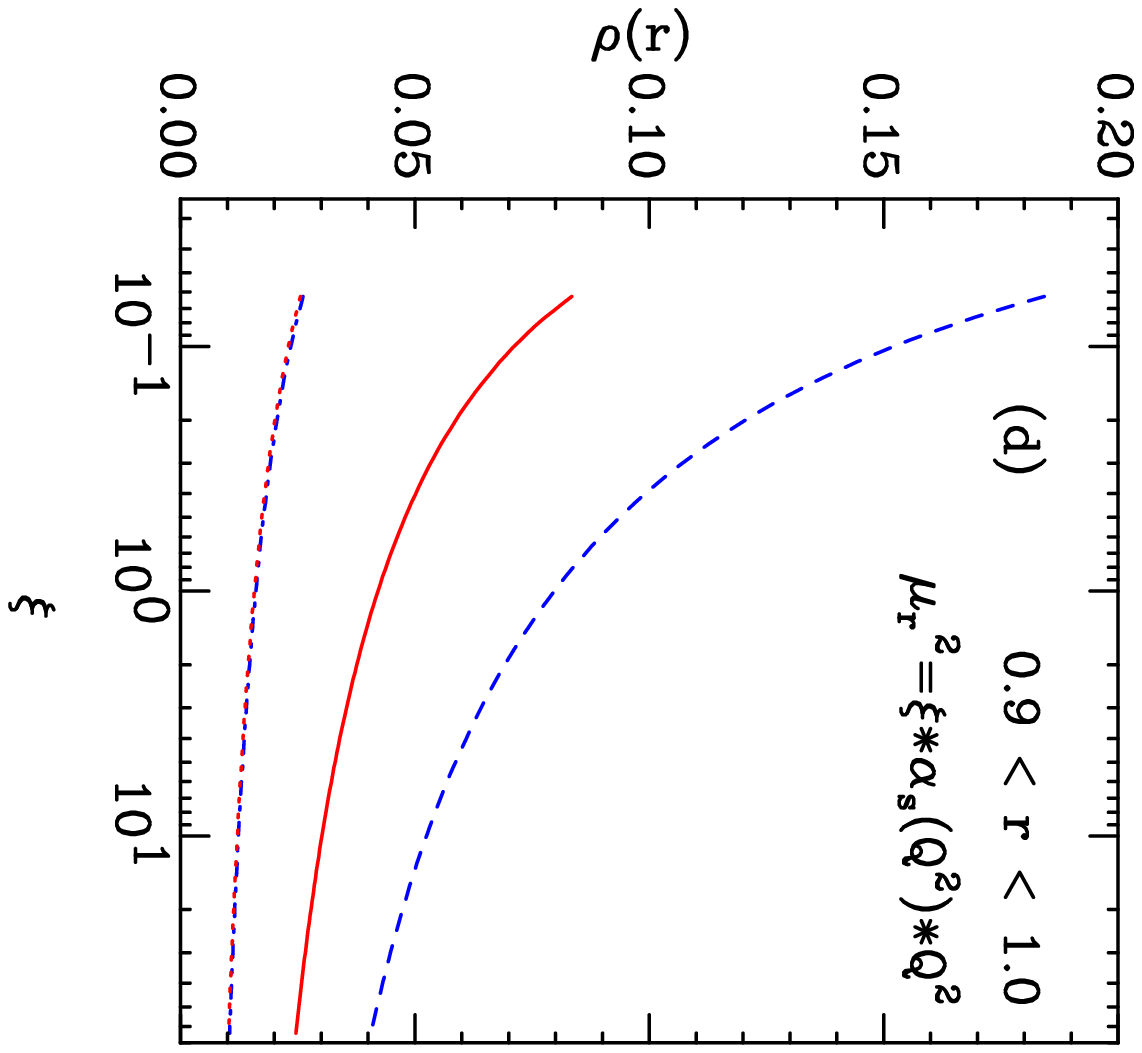}
\end{picture}
\vspace{11.0cm}
\caption{Renormalization scale dependence of the differential jet shape
$\rho(r)$ at (a) $r=0.15$, (b) $r=0.55$, (c) $r=0.75$, and (d) $r=0.95$,
for events satisfying the 1-jet cuts of 
Eqs.~(\protect\ref{eq:jetveto},\protect\ref{eq:cutsopt}).
Results are shown for the \Blue{PUCELL algorithm at LO (dash-dotted line) and
NLO (dashed line)} and for the \Red{$k_T$ algorithm at LO (dotted line) and NLO
(solid line)}. 
}
%\vspace*{0.2in}
\label{fig:scale.opt}
\end{figure}

The renormalization scale dependences of jet shapes in the PUCELL and 
the $k_T$ algorithm are compared in Fig.~\ref{fig:scale.opt}, using the 
1-jet cuts. Results are shown for four representative 
$r$-bins, at LO (dash-dotted and dotted lines) and at NLO (dashed and solid
lines). % for the PUCELL and the $k_T$-algorithm, respectively. 
The LO curves are virtually identical for the two algorithms. This
is to be expected, since the criterion for merging two partons is the same
at LO, namely their separation in the legoplot must be less than $R=1$. 
Small differences are either statistical or due to the different recombination
schemes for parton momenta. 
At NLO, the PUCELL and the $k_T$ algorithm produce quite 
similar results for the central part of the jet (at small $r$). In this 
region, sensitivity to the choice of renormalization scale is minimal 
near $\xi=1$ which confirms our basic choice of $\mu_r^2=\alpha_s Q^2$. 
At distances from the jet axis beyond $r\approx 0.5$, 
the iterative nature of the PUCELL algorithm is more likely
to gather a third parton into the jet, making it somewhat broader on average
than jets reconstructed by the $k_T$ algorithm. 
Since this broadening effect requires three partons
and is therefore only modeled at tree level in our NLO calculation, the 
NLO enhancement of $\rho(r)$ at large $r$ shows a pronounced scale 
dependence, which is significantly stronger in the PUCELL than in the $k_T$
algorithm. 

\begin{figure}[t]
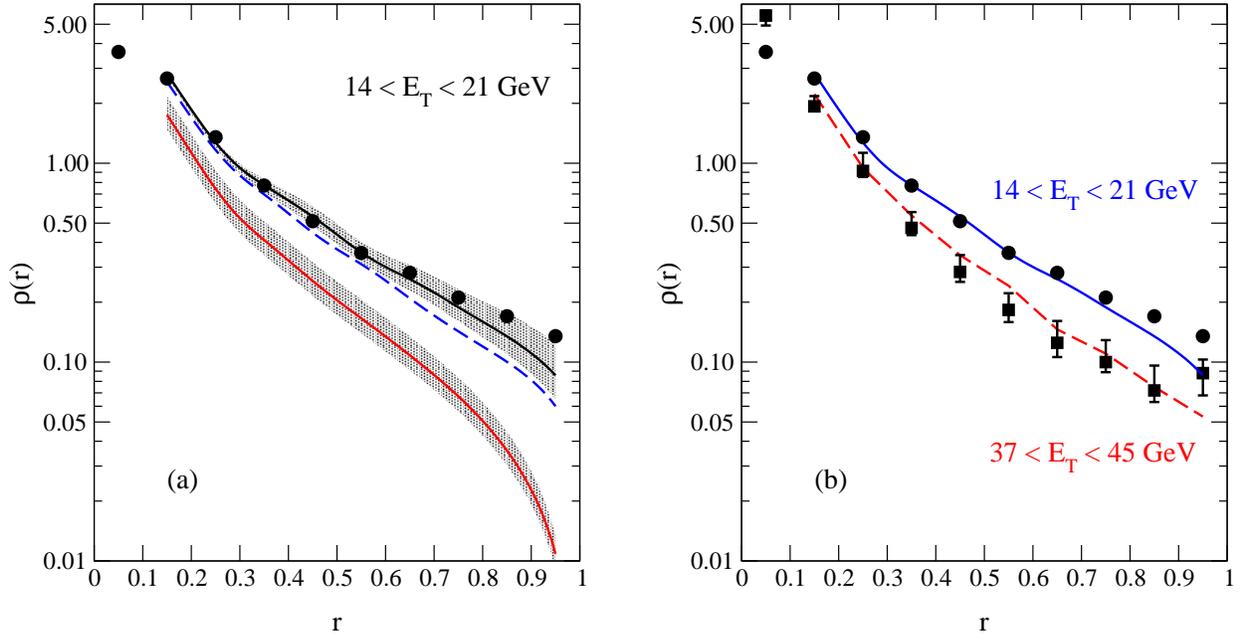

\vspace*{0.5in}
\begin{picture}(0,0)(0,0)
\includegraphics{fig4a.eps}
\includegraphics{fig4b.eps}
\end{picture}
\vspace{7.0cm}
\caption{Comparison of ZEUS jet shape data~\protect\cite{zeusdis} with 
QCD predictions for DIS jets reconstructed by the PUCELL algorithm. Jet 
cuts are: $-1<\eta<2$ and (a,b) 14~GeV~$<E_T<$~21~GeV,
(b) 37~GeV~$<E_T<$~45~GeV. ZEUS data (circles and squares) are 
compared in the lower $E_T$ range (a) with \Red{LO (lower band)} and 
\Blue{NLO (dashed line)} QCD predictions. The upper band in (a) and 
the two lines
in (b) represent  NLO jet shapes within the 1-jet cuts
of Eqs.~(\protect\ref{eq:jetveto},\protect\ref{eq:cutsopt}). 
The width of the bands corresponds to varying the 
renormalization scale between $\mu_r^2=\alpha_s Q^2/4$ and 
$\mu_r^2=4\alpha_s Q^2$.
}
\vspace*{0.2in}
\label{fig:zeus}
\end{figure}

Differential jet shapes in DIS, as a function of $E_T$ and pseudo-rapidity 
of the jet, have been measured by ZEUS~\cite{zeusdis}. 
In Fig.~\ref{fig:zeus}(a) we compare our LO and NLO QCD predictions with the 
ZEUS measurements for
events with at least one hard jet. This jet is identified with the PUCELL
algorithm and must lie in the interval 14~GeV~$<E_T<$~21~GeV and 
$-1<\eta<2$. The agreement between data and theory in the $0.1<r<1$ range
is improved significantly at NLO.
Calculation of the jet shape at $r=0$ is currently not possible, since it
would require the inclusion of two-loop
contributions and the resummation of multiple soft and collinear emission. 
As discussed previously, the scale dependence of the NLO results is 
quite large within the ZEUS acceptance cuts. The ZEUS data (and our NLO 
simulations) show, on the other hand, that the jet shapes depend
very little on the jet pseudo-rapidity, and modestly on jet $E_T$. Imposing
the 1-jet cuts of Eqs.~(\protect\ref{eq:jetveto},\protect\ref{eq:cutsopt})
does not significantly change $\rho(r)$ for the default scale choice of
$\mu_r^2=\alpha_s Q^2$, but it reduces the scale uncertainty. 
Included in Fig.~\ref{fig:zeus} are the NLO results using the 1-jet cuts.
The NLO QCD predictions
for $\rho(r)$ agree very well with the data, up to values around 
$r\approx 0.8$ for the 1-jet cuts and $r\approx 0.6$ for the generic ZEUS 
cuts. This corresponds to the regions where we have found a small
scale dependence. In other words, the differences between data and NLO
QCD are consistent with higher order QCD effects. Note that the agreement 
between theory and data is much worse at LO.

Similar agreement between true NLO QCD and ZEUS data is found for
different jet $E_T$ and $\eta$ ranges. One example is 
shown in Fig.~\ref{fig:zeus}(b), where jet shapes for jets with
14~GeV~$<E_T<$~21~GeV and 37~GeV~$<E_T<$~45~GeV are compared. 
The data and the NLO QCD predictions clearly show
that higher $E_T$ jets are narrower.
%, i.e., $\rho(r)$ decreases with
%increasing $E_T$, outside the jet center. 
It should be noted that the excellent agreement between data and theory, 
at the 10\% level, is obtained only when applying the minimal sensitivity 
criteria~\cite{stevenson} to pick the renormalization scale. A na\"{\i}ve 
choice, like $\mu_r=Q$ (roughly corresponding to $\xi\approx 10$ in 
Fig.~\ref{fig:scale.opt}) would lead to substantially larger deviations
from the experimental results.

We have performed a first analysis of true NLO jet shapes in DIS. We find 
excellent agreement between NLO QCD and data. This agreement is achieved 
without introducing extra phenomenological parameters describing hadronization
effects, like $R_{sep}$. For the PUCELL and the $k_T$ algorithm we have 
shown that the scale uncertainty of the QCD predictions for jet shapes is 
substantially reduced at NLO. A precise comparison of experiment with
theory is facilitated by cuts which suppress multi-jet events. 
Our analysis can be easily extended to other jet algorithms and 
momentum recombination schemes and allows one to investigate questions
like the infrared safety of jet algorithms or the reconstruction of kinematical
variables from jets at full NLO. Their effect on the matching of theory
and data for $ep$ collisions at HERA and more generally at hadron colliders 
can now be investigated at the one-loop level.

%
%%%%%%%%%%%%%%%%%%%%  ACKNOWLEDGMENTS  %%%%%%%%%%%%%%%%%%%%
%
%\newpage

\acknowledgements
This research was supported in part by the University of Wisconsin Research
Committee with funds granted by the Wisconsin Alumni Research Foundation and
in part by the U.~S.~Department of Energy under Contract
No.~DE-FG02-95ER40896.

%
%%%%%%%%%%%%%%%%%%%%%%%  REFERENCES  %%%%%%%%%%%%%%%%%%%%%%%
%

%\newpage

\end{document}